\documentclass[10pt]{iopart}
\usepackage{graphicx}

\begin{document}

\title[Author guidelines for IOPP journals]{Gravity Waves from Rotating 
Neutron Stars and Evaluation of Fast Chirp Transform Techniques}

\author{Tod E. Strohmayer\dag\ \footnote[3]{email: 
stroh@clarence.gsfc.nasa.gov}}

\address{\dag\ Laboratory for High Energy Astrohphysics, NASA/GSFC, Greenbelt, 
MD, 20771}

\begin{abstract}

X-ray observations suggest that neutron stars in low
mass X-ray binaries (LMXB) are rotating with frequencies
from 300 - 600 Hz. These spin rates are significantly less than the
break-up rates for essentially all realistic neutron star equations
of state, suggesting that some process may limit the spin 
frequencies of accreting neutron stars to this range. If the accretion induced 
spin up torque is in equilibrium with gravitational radiation losses, these
objects could be interesting sources of gravity waves.  
I present a brief summary of current measurements of neutron
star spins in LMXBs based on the observations of high-Q oscillations
during thermonuclear bursts (so called ``burst oscillations''). Further 
measurements of neutron star spins will be important in exploring the
gravitational radiation hypothesis in more detail. To this end I also present 
a study of fast chirp transform (FCT) techniques as described by Jenet
\& Prince (2000) in the context of searching for the chirping signals
observed during X-ray bursts.

\end{abstract}




\section{Introduction}

X-ray binaries are potentially among the most interesting sources of 
gravitational wave emission which current and future gravity wave detectors 
will attempt to study. The high frequency gravity wave signal produced during 
binary inspiral and ring down of black hole and neutron star binaries contains
detailed information on the properties of the compact object as well as the 
structure of spacetime in its vicinity. These objects will be prime targets 
for ground based detectors such as LIGO which because of seismic noise are 
only sensitive in the high frequency range above $\sim 100$ Hz.

Neutron stars are compelling targets of investigation because of the extreme 
physical conditions which exist in their interiors and immediate environs. 
For example, the gravity wave signals produced by inspiral of a neutron star
depend on the equation of state (EOS) at supranuclear density, a quantity 
which is still not precisely constrained by currently available astrophysics 
and nuclear physics data (see for example Heiselberg \& Hjorth-Jensen 1999). 
Moreover, fundamental properties of the star, such as its mass, can be 
extracted if the gravity wave signal can be measured. Thus gravity wave 
astronomy can in principle provide new probes of fundamental physics as well 
as advancing neutron star astrophysics. 

\section{Burst Oscillations}

Radio observations provided the first indications that some neutron stars 
are spinning with periods approaching 1.5 ms (Backer et al. 1982). These 
rapidly rotating neutron stars are observed as either isolated or binary 
radio pulsars. Binary evolution models indicate that neutron stars accreting 
mass from a companion can be spun up, or `recycled', to millisecond periods 
(see for example Webbink, Rappaport \& Savonije). 
This formation mechanism likely accounts for a substantial fraction of the 
observed population of millisecond radio pulsars, however, other formation 
scenarios have also been proposed (see van den Heuvel \& Bitzaraki 1995; 
van Paradijs et al. 1997).

\begin{table}[tbp]
\begin{center}
\centerline{{\bf Table 1.} Burst Oscillation Sources and Properties}
\begin{tabular}{cccc}
\hline
Sources & Frequency (Hz) & $\Delta\nu$ (kHz QPO, in Hz) & 
References$^1$ \cr
\hline
4U 1728-34 & 363 & 363 - 280 & 1, 2, 3, 4, 5, 13, 14 \cr
4U 1636-53 & 290, 580 & 251 & 6, 7 \cr
4U 1702-429 & 330 & 315 - 344 & 4, 9 \cr
KS 1731-260 & 524 & 260 & 10, 11, 12 \cr
Galactic Center & 589 & Unknown & 15 \cr
Aql X-1 & 549 & Unknown & 16, 17 \cr
X1658-298 & 567 & Unknown & 18 \cr
4U 1916-053 & 270 & 290 - 348 & 19, 20 \cr
4U 1608-52 & 619 & 225 - 325 & 8, 21 \cr
SAX J1808-369 & 401 & Unknown & 22,23 \cr
\hline
\end{tabular}
\end{center}
\hfil\hspace{\fill}

{\small $^1$References: (1) Strohmayer et al. (1996); (2) Strohmayer,
Zhang, \& Swank (1997); (3) Mendez \& van der Klis (1999); (4) Strohmayer 
\& Markwardt (1999); (5) Strohmayer et al. (1998b); (6) Strohmayer et al. 
(1998a); (7) Miller (1999); (8) Mendez et al. (1998); (9) Markwardt, 
Strohmayer
\& Swank (1999) (10) Smith, Morgan, \& Bradt (1997); (11) Wijnands \& van der
Klis (1998); (12) Muno et al. (2000); (13) van Straaten et al. (2000); (14)
Franco (2000); (15) Strohmayer et al (1997); (16) Zhang et al. (1998); (17)
Ford (1999); (18) Wijnands, Strohmayer \& Franco (2000); (19) Boirin et al. 
(2000); (20) Galloway et al. (2000); (21) Chakrabarty (2000); (22) Heise 
(2000); (23) Ford (2000)}
\end{table} 

In recent years direct evidence linking the formation of rapidly rotating 
neutron stars to accreting X-ray binaries has been provided by data from the 
Rossi X-ray Timing Explorer (RXTE). The first evidence came from the discovery
of high amplitude, nearly coherent X-ray brightness oscillations (so called 
`burst oscillations') during thermonuclear flashes from several neutron star 
binaries (see Strohmayer 2000 for a recent review). These oscillations likely 
result from spin modulation of either one or a pair of antipodal `hot spots' 
generated as a result of the thermonuclear burning of matter accreted on the 
neutron star surface. Indisputable evidence that neutron stars in X-ray 
binaries can indeed be rotating rapidly then came with the discovery of the 
first accreting millisecond X-ray pulsar SAX J1808-369 (Wijnands \& van der 
Klis 1998; Chakrabarty \& Morgan 1998), which is spinning at 401 Hz. 

The observed distribution of burst oscillation
frequencies, including the 401 Hz pulsar, is very similar to the observed 
distribution of millisecond radio pulsars. Table 1 lists properties of
burst oscillations in the sources in which they have been detected. 
The RXTE observations suggest that the spin frequencies of neutron stars in 
accreting binaries span a relatively narrow range from 300 - 600 Hz. Moreover, 
these observed frequencies are significantly less than the maximum neutron 
star spin rates for almost all but the stiffest neutron star equations of 
state (Cook, Shapiro \& Teukolsky 1994). In principle, accretion should drive
the spin frequencies close to break-up, unless some other mechanism intervenes
to remove the accreted angular momentum.

\begin{figure}
\centering
 \includegraphics[width=4in,height=3in]{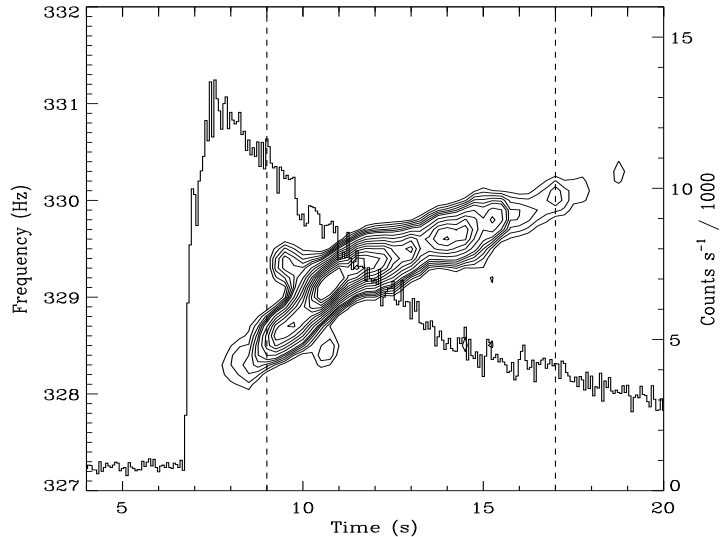}
\caption{Contour plot of the dynamic power spectrum of a burst from the LMXB 
4U 1702-429 observed on April 1, 2001 (UTC) with the PCA onboard RXTE. Shown 
are contours of constant power spectral amplitude as a 
function of frequency and time. The burst profile is also plotted (right axis).
The oscillation frequency increase from about 328 to 330 Hz in $\sim 10$ 
seconds.}
\end{figure}

\section{Gravity Wave Emission from Rotating Neutron Stars}

Bildsten (1998) recently proposed that the angular momentum gain from accretion
could be offset by gravitational radiation losses if a misaligned quadrupole 
moment of order $10^{-8} - 10^{-7} I_{NS}$ could be sustained in the neutron 
star crust. Here $I_{NS}$ is the moment of inertia of the neutron star. In 
this scenario the strong spin frequency dependence of gravitational radiation 
losses sets the limit on the observed spin frequencies.  Recent theoretical 
work has also shown that an $r$-mode instability in rotating neutron stars may 
also be important in limiting the spins of neutron stars via gravity wave 
emission (see Ushomirsky, Bildsten \& Cutler 2000; Bildsten 1998; Levin 1999; 
Andersson, Kokkotas, \& Stergioulas 1999).

The strong $\nu^6$ frequency dependence of the energy radiated by gravitational
waves means that rapidly rotating neutron stars with misaligned quadrupole 
moments might have observationally interesting gravitational wave amplitudes. 
The spin periods of neutron stars inferred from burst oscillations cluster 
rather tightly in the range from 300 - 600 Hz. White \& Zhang (1997) 
suggested that the observed range of spin frequencies could be produced if 
these neutron stars were spinning in magnetic equilibrium. However, in order 
for the observed frequencies to be similar would require that mass accretion 
rate and the magnetic moment be correlated. It is not presently known if such 
a correlation is to be expected based on theoretical grounds. 
The model proposed by Bildsten (1998) suggests that 
the spins of these neutron stars may be limited by the emission of gravity 
waves (see also Wagoner 1984). The spin down torque due to gravitational wave 
emission is proportional to $\nu^5$ so that one would expect a critical spin 
frequency above which accretion torques are cancelled out by gravity wave 
losses. By equating the characteristic accretion torque with the gravity wave 
torque one can determine the average quadrupole moment required to maintain 
the critical frequency. For the mass accretion rates characteristic of LMXBs 
and a critical spin frequency of 300 Hz one obtains a quadrupole $Q \sim 4.5 
\times 10^{37}$ gm cm$^2$, or about $5 \times 10^{-8} I_{NS}$ (see Bildsten 
1998). The question that remains is whether or not such a quadrupole moment 
can be routinely generated within a neutron star. This is presently an area of
active research (see for example, Ushomirsky, Bildsten \& Cutler 2000).

Regardless of the mechanism, if the accretion torque is indeed balanced by 
gravity wave losses then the amplitude of the gravitational radiation can be 
calculated (see Wagoner 1984; Bildsten 1998). The dimensionless strain $h$ is 
in the range from $h \sim 10^{-27}  - 10^{-26}$. Although less than the 
estimated sensitivity for LIGO I, one can greatly improve the sensitivity by 
pulse folding if the rotational ephemeris of the neutron star is known (see 
Brady \& Creighton 1999; Ushomirsky, Bildsten \& Cutler 2000). Current 
estimates indicate that a narrow band configuration for LIGO-II will reach 
interesting search limits for these neutron stars, especially for the brightest
of the LMXBs, for example, Sco X-1 (Ushomirsky, Bildsten \& Cutler 2000). This 
also provides strong motivation for additional deep X-ray timing searches in 
order to detect coherent pulsations in more LMXBs and to measure the pulse 
ephemerides so as to improve searches for gravity wave emission. It also 
illustrates the strong synergism between X-ray and gravity wave astronomy in 
the context of neutron star binaries.

\section{Searches for Chirping Signals: Burst Oscillations}

One way to search for additional neutron star spin periods is to try and 
make more sensitive searches for burst oscillations. Since the signals during
many bursts are known to increase with time (ie. they are chirps, see 
Figure 1), it is likely that more general search techniques which allow for 
frequency changes will be better at detecting weak burst oscillation signals. 
Prince \& Jenet (2000) have recently described a general method based on 
multidimensional FFTs to search for signals with varying frequency. They call 
this technique the Fast Chirp Transform (FCT). In an effort to develop 
more sensitive searches for burst oscillations as well as techniques to 
search for GW emission from rotating neutron stars I have begun 
evaluating the FCT on X-ray burst oscillation data which are known to contain 
chirping signals.  Here we briefly describe the simplest 2-parameter (linear) 
chirp and the corresponding 2-d FCT.

A signal with a linearly increasing frequency is of the form $h_s(t) = A(t) 
\sin (2\pi (\nu t + \dot\nu t^2))$, that is the phase advance is 
quadratic in time. The FCT is an approximation to the exact Fourier transform 
for such a signal; $H_{k_0,k_1} = \sum_{j_0}  h_{j_0} \exp (2\pi i 
(k_0(j_0/N_0) + k_1(j_0/N_0)^2))$. The second term in the exponential 
represents the quadratic phase advance produced by the changing frequency. 
The sum is over $j_0$ with $0 <  j_0 < N_0 - 1$. That is, the data, 
$h_{j_0}$ , are a discretely sampled time series with $N_0$ bins. The FCT 
approximates this expression by breaking up the single sum into a double sum 
by defining a set of $N_1$ subintervals of the data. Within each subinterval
the phase advance produced by the quadratic term is required to have the same 
value, that is, it increments uniformly as a ``second index'', and the 
approximate expression can be evaluated as a 2-d FFT. This requires only 
$O (N_0N_1 log_2(N_0 N_1))$ operations as opposed to $O (N_0^2)$, and is 
therefore much more computationally efficient. There is a simple geometrical
way to see how the 2-d FFT implementation of the FCT is constructed. One forms
a two dimensional array, one dimension being $N_0$ in size (the number of 
samples in the original time series), and the other being of length $N_1$ 
(the number of subintervals). The length of each of the $N_1$ data segments 
packed along the $N_0$ component is determined by the phase evolution function.
Since the increase in phase is quadratic in time, subsequent data segments 
down the $N_1$ direction become shorter because they must represent the same 
total phase advance.  In this way the boundaries of the subintervals are
specified uniquely by the phase evolution function (see Prince \& Jenet 2000). 
 
The range of $\dot\nu$ searched is determined by $N_1$, the number of 
subintervals of the original time series. The maximum $\dot\nu$ searched
is simply $\dot\nu_{max} = N_1 / 2T^2$, where $T$ is the length of the 
original time series. Both positive and negative values are searched, 
analogously to the positive and negative frequencies of a standard FFT
power spectrum, in this case, however, the FCT is not in general symmetric
with regard to positive or negative $\dot\nu$. 

I have implemented the 2-d FCT in IDL using the formalism outlined in Jenet 
\& Prince (2000). I have evaluated the FCT by looking at
burst oscillation data in bursts with detected oscillations from the LMXB 
4U 1702-429 (see Markwardt, Strohmayer \& Swank 1998). Figure 2 shows a 
comparison between a standard FFT power spectral analysis and
one using the FCT for the burst shown in Figure 1. As can be see in Figure 1, 
this burst has a strong oscillation at $\sim 330$ Hz. I used 8 seconds of data 
sampled at 4096 Hz (the interval between the dashed vertical lines in 
Figure 1). The data were first mixed with a $f_{mix} = 326$ Hz modulation in 
order to bring the oscillation signal down to lower frequency, and therefore
reduce $N_0$. This is similar to a heterodyne procedure. Figure 2 (left) shows 
a plot of the peak power in the FCT as a function of $\dot\nu$. One can 
see that the signal strength is greatly increased at $\dot\nu \sim
0.075$ s$^{-2}$ compared with $\dot\nu = 0$. In Figure 2 (right) the power 
spectrum with the highest value of the FCT power is compared with that computed
with $\dot\nu = 0$ in Figure 2 (right). The FCT at positive $\dot\nu$ recovers 
a narrow, coherent peak whereas that computed with no frequency derivative is 
smeared out and weaker.  This demonstrates the effectiveness of the FCT for 
detecting these kinds of signals. A detailed search for burst oscillations in
new data will be presented elsewhere. 

\ack I thank Tom Prince for many helpful discussions and suggestions. 

\begin{figure}
\centering
  \begin{minipage}[c]{0.5\textwidth}
    \centering \includegraphics[width=2.8in, height=2.5in]{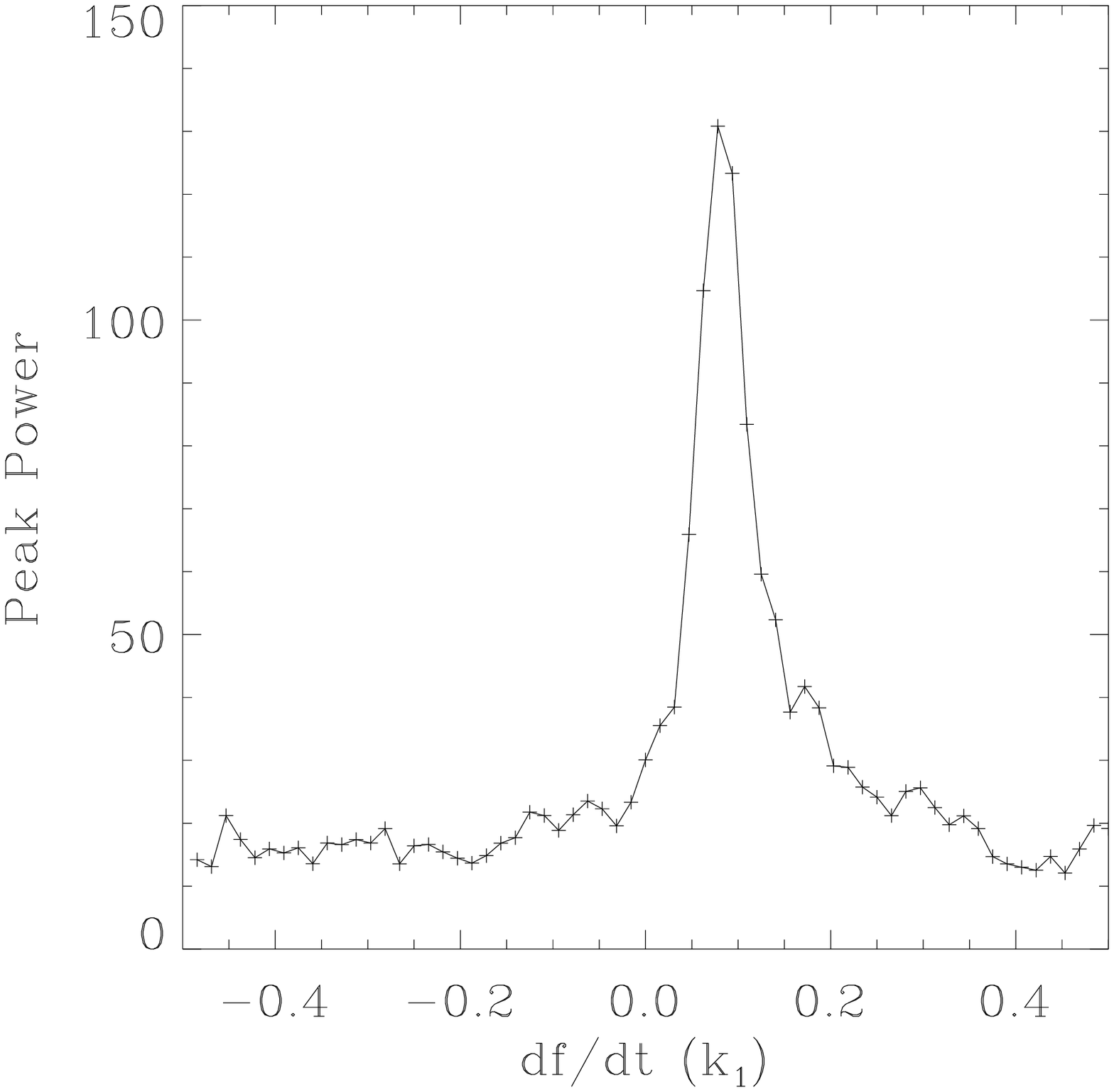}%
  \end{minipage}%
  \begin{minipage}[c]{0.5\textwidth}
    \centering \includegraphics[width=2.8in, height=2.5in]{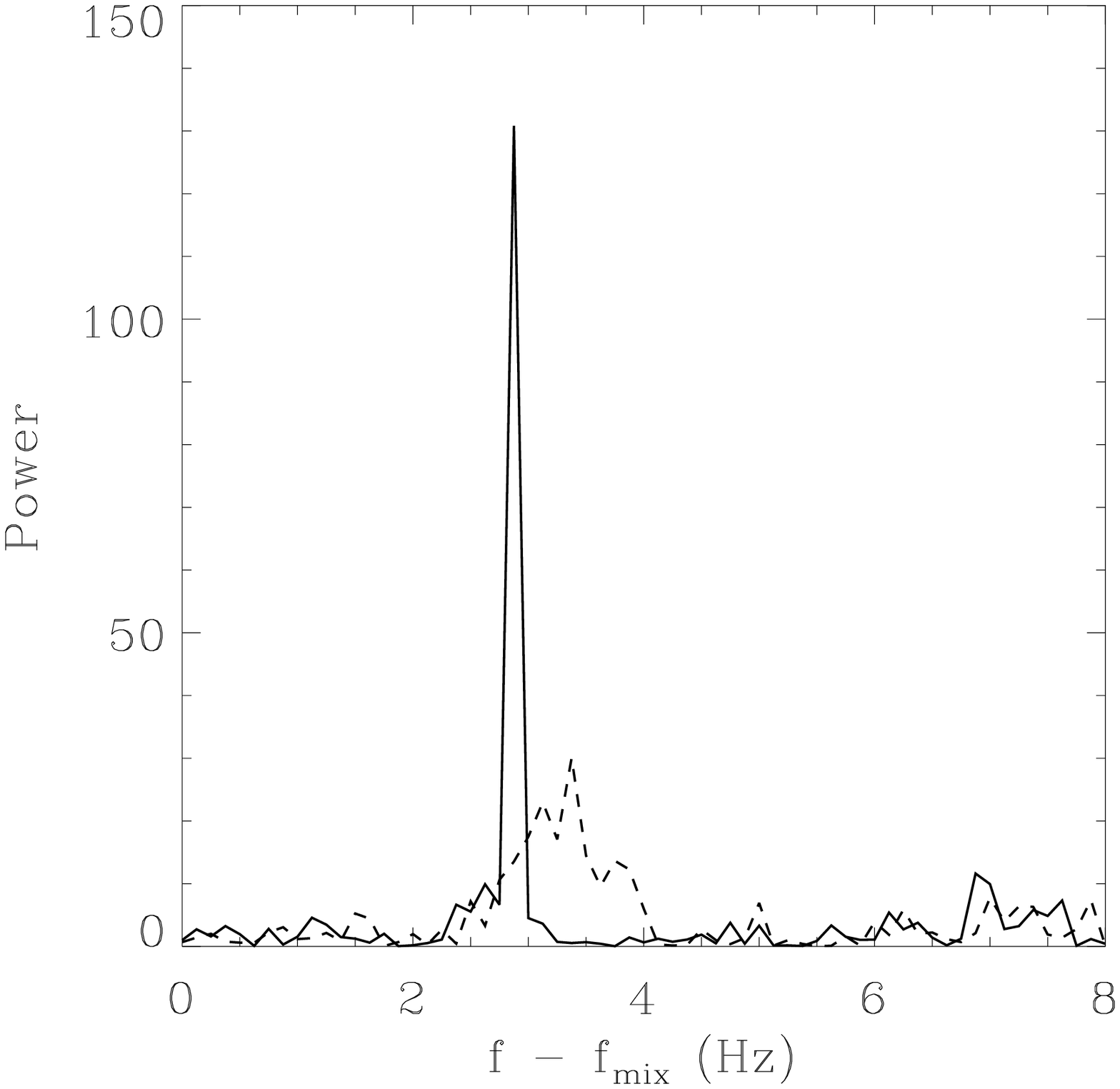}
  \end{minipage}
\centering
\caption{Fast chirp transform (FCT) analysis of a burst from 4U 1702-429. 
The left panel shows the peak FCT power as a funtion of frequency derivative,
$\dot\nu$. The right panel compares the FCT power spectra at two different 
values of $\dot\nu$, $\dot\nu=0$ (dashed) and $\dot\nu=0.078$ (solid).}
\end{figure}

\References

\item[] Andersson, N. Kokkotas, K. D. \& Stergioulas, N. 
1999, ApJ, 516, 307

\item[] Backer, D. C., Kulkarni, S. R., Heiles, C., Davis, M. M. \& Goss, W. 
     M. 1982, Nature, 300, 615

\item[] Bildsten, L. 1998, ApJ, 501, L89

\item[] Boirin, L., et al. 2000, A\& A, 361,121

\item[] Brady, P. \& Creighton, T. 1999, Phys. Rev. D, in press 
(gr-qc/9812014)

\item[] Chakrabarty, D. 2000, Talk presented at AAS HEAD meeting, 
     Honolulu, HI

\item[] Chakrabarty, D. \& Morgan, E. H. 1998, Nature, 394, 346

\item[] Ford, E. C. 2000, ApJ, 535, L119

\item[] Ford, E. C. 1999, ApJ, 519, L73

\item[] Franco, L. 2000, ApJ, submitted (astro-ph/0009189)

\item[] Galloway et al. 2000, ApJ, submitted (astro-ph/0010072)

\item[] Heise, J. et al. 2000, Talk presented at AAS HEAD meeting, 
     Honolulu, HI

\item[] Heiselberg, H.\ \& Hjorth-Jensen, M.\ 1999, ApJ, 525, L45

\item[] Levin, Y. 1999, ApJ, 517, 328

\item[] Markwardt, C. B., Strohmayer, T. E., \& Swank, J. H. 1999, ApJ, 512, 
L125

\item[] Mendez, M. \& van der Klis, M. 1999, ApJ, 517, L51

\item[] Mendez, M., van der Klis, M. \& van Paradijs, J. 1998, ApJ, 506, L117

\item[] Miller, M. C. 1999, ApJ, 515, L77

\item[] Muno, M.\ P., Fox, D.\ W., Morgan, E.\ H.\ \& Bildsten, L.\ 2000, 
ApJ, 542, 1016

\item[] Prince, T. A. \& Jenet, F. A. 2000, Phys. Rev. D, 62, 122001

\item[] Smith, D., Morgan, E. H. \& Bradt, H. V. 1997, ApJ, 479, L137

\item[] Strohmayer, T. E. et al. 1996, ApJ, 469, L9

\item[] Strohmayer, T. E., Zhang, W. \& Swank, J. H. 1997, ApJ, 487, L77

\item[] Strohmayer, T. E., Jahoda, K., Giles, A. B. \& Lee, U. 1997, ApJ, 
486, 355

\item[] Strohmayer, T. E. \& Markwardt, C. B. 1999, ApJ, 516, L81

\item[] Strohmayer, T. E., Zhang, W., Swank, J. H. \& Lapidus, I. 1998b, 
ApJ, 503, L147

\item[] Strohmayer, T. E., Zhang, W., Swank, J. H., White, N. E. \& 
Lapidus, I. 1998a, ApJ, 498, L135

\item[] Strohmayer, T. E. 2000, in Proceedings of X-ray Astronomy '99,
Stellar Endpoints, AGN and the Diffuse X-ray Background. Bologna, Italy, 
(astro-ph/9911338)

\item[] Ushomirsky, G., Bildsten, L. \& Cutler, C. 2000, in 3rd Edoardo
Amaldi Conference on Gravitational Waves, (astro-ph/0001129)

\item[] van den Heuvel, E.\ P.\ J.\ \& Bitzaraki, O.\ 1995, A\& A, 297, L41

\item[] van Paradijs, J., van den Heuvel, E.\ P.\ J., Kouveliotou, C., 
Fishman, G.\ J., Finger, M.\ H.\ \& Lewin, W.\ H.\ G.\ 1997, A\& A, 317, L9

\item[] van Straaten, S. et al. 2000, ApJ, submitted, (astro-ph/0009194)

\item[] Wagoner, R. V. 1984, ApJ, 278, 345

\item[] Webbink, R. F., Rappaport, S. A. \& Savonije, G. J. 1983, ApJ, 
270, 678

\item[] White, N. E. \& Zhang, W. 1997, ApJ, 490, L87 

\item[] Wijnands, R. \& van der Klis, M. 1998, Nature, 394, 344

\item[] Wijnands, R. Strohmayer, T. E. \& Franco, L. M. 2000, ApJ, in press 
     (astro-ph/0008526)

\item[] Zhang, W. et al. 1998, ApJ, 495, L9-12

\endrefs

\end{document}